    \documentclass[12pt,psf,epsf]{article}
\usepackage[centertags]{amsmath}
\usepackage{amssymb}
\usepackage{graphicx}
\usepackage{epsfig}
\usepackage{ulem}
\usepackage[english]{babel}
\usepackage{array}
\usepackage{amsthm}
\usepackage{latexsym}
\usepackage[mathcal]{euscript}
\pdfoutput=1
\usepackage{epsfig}

 \usepackage{jheppub}
 \usepackage{hyperref}

\usepackage{subfigure}
\usepackage{psfrag}

\usepackage{epsfig}
\usepackage[latin1]{inputenc}
\usepackage{float}
\usepackage{graphicx}
\usepackage{cancel}
\usepackage{mathrsfs}
\usepackage{amssymb}
\usepackage{amsfonts}
\usepackage{amsmath}
\usepackage{slashed}
\usepackage{graphicx}
\usepackage{bm}
\usepackage{color}
\newcommand{\be}{\begin{equation}}
\newcommand{\ee}{\end{equation}}
\newcommand{\bea}{\begin{eqnarray}}
\newcommand{\eea}{\end{eqnarray}}
\newcommand{\bwt}{\begin{widetext}}
\newcommand{\ewt}{\end{widetext}}

\newcommand{\nn}{\nonumber}

\newcommand{\bi}{\begin{itemize}}
\newcommand{\ei}{\end{itemize}}

\newcommand\sE{{\ensuremath{{\mathcal E}}}}

\newcommand{\ff}{\mathfrak{f}}

\newcommand{\fg}{\mathfrak{g}}
\newcommand{\fn}{\mathfrak{n}}
\newcommand{\fk}{\mathfrak{k}}

\newcommand{\fr}{\mathfrak{r}}
\newcommand{\fq}{\mathfrak{q}}
\newcommand{\fx}{\mathfrak{x}}
\newcommand{\fv}{\mathfrak{v}}

\newcommand{\fS}{\mathfrak{S}}
\newcommand{\fK}{\mathfrak{K}}
\newcommand{\fN}{\mathfrak{N}}

\newcommand{\fM}{\mathfrak{M}}

\usepackage{setspace}
\begin{document}

\title {Holographic Non-equilibrium Heating}

\author{D.S.Ageev and I.Ya.Aref'eva}
\affiliation{Steklov Mathematical Institute, Russian Academy of Science}
\emailAdd{ageev@mi.ras.ru}
\emailAdd{arefeva@mi.ras.ru}

\abstract{
We study the holographic entanglement entropy evolution after a global sharp quench of thermal state. 
  After the quench, the system comes to equilibrium and the temperature increases 
from $T_i$ to $T_f$. Holographic dual of this process is provided by an injection of a thin shell of matter in the black hole background. 
The quantitative characteristics of the evolution  depend substantially on the size of the initial black hole.
 We show that characteristic  regimes
during non-equilibrium heating  do not depend on the initial temperature and are the same as in thermalization. 
Namely these regimes are pre-local-equilibration quadratic growth,  linear growth and  saturation  regimes  of the time evolution of the holographic entanglement entropy. We study the initial temperature dependence of quantitative characteristics of these  regimes and find that the critical exponents do not depend on the temperature,  
meanwhile the prefactors are the functions  on the temperature. 
}

\maketitle

\newpage

\section{Introduction}
 It is universally recognized that one of the most challenging problems  in  quantum theory is the  description of the equilibration process, in particular, the thermalization. The AdS/CFT correspondence \cite{Malda,GKP,Witten} proposes a powerful tool for a description of thermalization process in general class of strongly coupling quantum  theories.  In the holographic duality, the temperature in the quantum field theory is  related with the black hole (or black brane) temperature  in the dual background \cite{WittenTH}-\cite{MaldacenaET}. The thermalization   within AdS/CFT duality corresponds to
 a black hole formation, see  \cite{CasalderreySolana:2011us,IA,DeWolf,Hartnoll:2016apf,Easther:2011wh}.   The simplest description of a black hole formation process is provided by the Vaidya deformation of a given background.
 This model serves as a relatively simple and universal holographic model of thermalization with a wide range of applicability in different physical situations, see \cite{Balasubramanian:2011ur}-\cite{1602.05934} and references therein. Thick shell Vaidya models can be solved only numerically, meanwhile several thin shell models admit an analytical solution \cite{Balasubramanian:2011ur,Lopez,ABK}.  Thin shell models capture all  features of evolution during sharp quenches \cite{HL,HLlong} and in the low-dimensional case reproduce results obtained in the  conformal field theory  \cite{Calabrese:2005in,Calabrese:2006}.
 
  All these models have zero initial temperature.
However, not all interesting non-equilibrium processes in strongly correlated systems start from zero temperature.
  Especially, this concerns phenomena related to biological systems. 
   In particular, application of  holographic approach  to  a part of the  photosynthetic process \cite{AV-photo} requires consideration of  the global quench of thermal states. From non-holographic point of view, a global quench in quantum field theory starting from a thermal initial state has been considered for massive non-interacting field models in \cite{QQT}.

In this paper, in the holographic approach, we study the simplest process associated with a global quench of the thermal initial state, which leads to an increase in temperature.
As a holographic dual to this process,
we use the so-called double-BTZ-Vaidya background.  This background interpolates between two AdS black holes with different temperatures. The thin shell limit of this background corresponds to the sharp quench.
We show that holographic non-equilibrium heating  inherits  typical properties of holographic thermalization.  
Behaviour of the holographic entanglement entropy (HEE) during thermalization  have been studied in details by H.Liu and J.Suh \cite{HL,HLlong}, see also \cite{Hubeny:2013hz, Li:2013sia,Shenker:2013pqa}. They have shown, that during thermalization the HEE passes  several regimes, namely, the pre-local-equilibration quadratic growth, the post-local-equilibration linear growth, the  late-time regime, and the saturation regime. We show that all these regimes  are present in  non-equilibrium heating.
We also show that although the evolution of the large regions of entanglement entropy 
is controlled by the geometry around and within the horizon  of the emerging black hole,  the quantitative characteristics of  evolution from the thermal state also depend substantially on the horizon of the initial black hole.
We calculate the corrections to different scaling coefficients taking into account non-zero initial temperature.
 
The paper is organized as follows. In Section \ref{Sect:DG} we sketch the dual geometry of  non-equilibrium heating process. Then in Section \ref{Sect:Formula} we present the new explicit formulae 
for the entanglement entropy evolution after the global quench of initially thermal state.
After that, in  Section \ref{Sect:critline}, we study in details  relations between  bulk characteristics of the geodesics anchored on the interval and the boundary data.  In Section \ref{Sect:MLR1} we list the specific regimes in non-equlibrium heating and compare with some previous results about behavior of the entanglement entropy  during  thermalization. Section \ref{quadrgrowth} is devoted to the initial quadratic growth regime. In Section \ref{MLR3}  we compute the scaling characteristics of the memory loss regime in the  process of the global quench of thermal state with temperature $T_i\neq 0$. The main focus here is on their  dependences on the final and initial temperatures.  In the final Section \ref{Sect:Final} we conclude and we discuss some possible generalizations.

\newpage

\section{Holographic entanglement entropy in dBTZ-Vaidya background}
\label{Sect:setup}

\begin{figure}[h!]
\centering
\begin{picture}(185,35)
\put(30,0){singularity}
\put(120,0){singularity}
\put(20,-70){AdS}
\put(44,-17){BH $z_h$}
\put(163,-17){BH $z_h$}
\put(120,-23){BH $z_H$}
\put(85,-95){$z=0$}
\put(27,-35){$v=0$}
\put(155,-45){$v=0$}
\end{picture}\\$\,$
	\put(-120,0){\includegraphics[width=4cm]{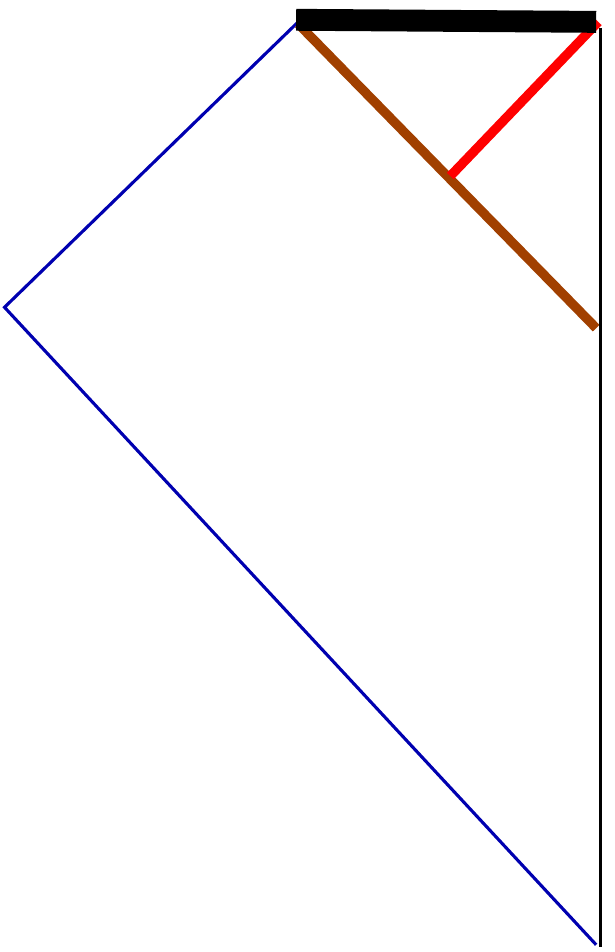}}
\put(0,63){\includegraphics[width=4cm]{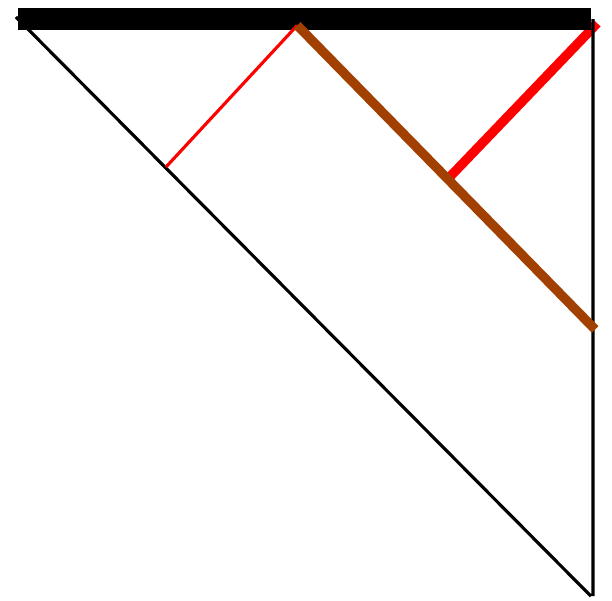}}
 \caption{Penrose diagrams for double-BTZ-Vaidya geometry defined by \eqref{vaidyaT} and \eqref{BHpm}. The left plot corresponds to $z_H=0$ (usual BTZ-Vaidya spacetime), the right plot corresponds to the case $z_H>0$.}
 \label{fig:FF}
\end{figure}
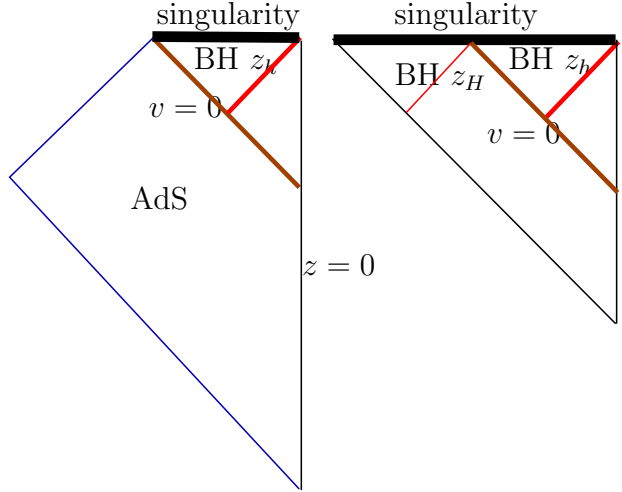

 \subsection{The dual geometry}\label{Sect:DG}
 We are interested in the evolution of holographic entanglement entropy of the interval of length $\ell$ and at time $\tau$ in holographic setup described by the double-BTZ-Vaidya quench (see Fig.\ref{fig:FF}). We call double-BTZ-Vaidya  or dBTZ-Vaidya  the geometry that describes shell of null matter  accreting in the black hole background. We focus on the thin shell limit of this metric. In contrast to usual BTZ-Vaidya geometry (sometimes it is called AdS-Vaidya geometry) black hole is present from the very beginning of process.
Thus we consider black hole evolving from the initial state defined by the horizon position $z_H$ to the final state with horizon $z_{h}$ as a dual background. 
    The dBTZ-Vaidya metric in the thin shell limit is given by
\bea\label{vaidyaT}
ds^2&=& {R^2 \over z^2} \left(- f (z,v) dv^2 - 2 dvdz + d \vec x^2 \right),\,\,\,\,\,\,f(z,v)=\theta(v)f_{h}(z)+\theta(-v)f_H(z)
\eea
and functions  $f_{H}$ and $f_{h}$ are defined as
\bea\label{BHpm}
f_H=1-\left(\frac{z}{z_{H}}\right)^2,\,\,\,\,
f_{h}=1-\left(\frac{z}{z_{h}}\right)^2,\,\,\,\,z_h<z_{H},
\eea
where the time is related with variable $v$ as
\be\label{Vm1}
v<0:\,\,\,
\,\,\,\,t=v+z_H \text{arctanh} \frac{z}{z_H};\,\,\,\,\,\,\,\,
v>0:
\,\,\,
\,\,\,\,t=v+z_h \text{arctanh} \frac{z}{z_h}.
\ee
Usual holographic BTZ-Vaidya setup is recovered in the limit $z_H\rightarrow\infty$. 
The initial and final temperatures and energy densities are
\be
\label{temperatures}
T_i=\frac{1}{2\pi z_H},\,\,\,\,\,\,\,\,T_f=\frac{1}{2\pi z_h},\,\,\,\,\,\,\,\,
\sE_i = {R\over 8 \pi G_N}  \frac{1}{ z^2_H},\,\,\,\,\,\,\,\,\sE_f = {R\over 8 \pi G_N}  \frac{1}{ z^2_h},\ee
 where $G_N$ is Newton's constant in the bulk and $R$ is a typical scale in the bulk. 
 Note, that the case $z_H<z_h$ corresponds to a model of cooling \cite{IAIV} and this model violates NEC condition \cite{HL}.

\subsection{The geodesic length}\label{Sect:Formula}
Now let us turn to the description of the  holographic entanglement entropy of the interval in non-equilibrium heating process dual to the metric \eqref{vaidyaT}. This entanglement entropy is obtained explicitly as  a length of the  geodesic that are spacelike, with both endpoints anchored on the boundary at the same time $\tau$ and  length separation $\ell$. Also these geodesics are called ETEBA (Equal-Time Endpoints Boundary Anchored) geodesics for brevity, see \cite{1312.6887}. In this paper we set $R=1$ and also ignore the factors of $4G$ identifying the length of ETEBA geodesics with the entanglement entropy.
  This class of geodesics in dBTZ geometry has been considered in \cite{DAIA}. For the sake of convenience we introduce parameter $\kappa=z_h/z_H$. The formula for entanglement entropy has the form
 \be 
 \label{S-AA}S= \log  \left( \frac{R}{\varepsilon} \frac{z_h} {\ell \,\fS_\kappa (\rho,s)}\,\sinh\frac {\tau } {z_h}\right),
  \ee
  where the form of $\fS_\kappa$ will be given below (see formula \eqref{fS}). This formula has the form similar to that from \cite{Balasubramanian:2010ce} and in the limit $\kappa\rightarrow0$ the function $\fS_0(\rho,s)=s$ recovering the result of \cite{Balasubramanian:2010ce}. Formula \eqref{S-AA} contains parameters $\rho$ and $s$ related with the bulk characteristics of the geodesic
\bea
s=\frac{z_c}{z_*},\,\,\rho=\frac{z_h}{z_c},\,\,\,c=\sqrt{1-s^2},\,\,\, s=\sin \phi,
\eea 

  where $z_c$ is the point where the geodesic crosses the shell and $z_*$ is the turning point of the geodesic. Using these parameters the explicit formula for $\fS_\kappa$ has the form
  \be\label{fS}
  \fS_\kappa (\rho,s)=\frac {c \rho + \Delta} {\Delta}\cdot\sqrt {\frac {\Delta^2 - 
          c^2 \rho^2} {\rho\left (c^2 \rho + 
             2 c\Delta + \rho \right) - \kappa^2}},\ee
and for simplicity we define $\gamma=1-\kappa^2$ and $\Delta=\sqrt{\rho^2-\kappa^2}$.
The final ingredient  to describe the geodesic length is the relation obtained in \cite{DAIA} between boundary separation $\ell$, time $\tau$ and bulk data $\rho$ and $s$. These formulae, being quite complicated are the explicit generalization of the similar relation from \cite{Balasubramanian:2010ce}. Let us define the total length $\ell$ as $\ell=\ell_-+\ell_+$.  Here  $\ell_-$ is the length of the part of the geodesic under the shell, i.e. between points  $z_*$ and $z_c$, while $\ell_-$ is the length of the part of the geodesic connecting points $z_c$ and boundary $z=0$ at $x=\ell$  over the shell. The expressions for $\tau$ and $\ell$ take the form
 \bea\label{t}
&&\frac{\tau}{z_h}={\mbox {arccoth}}\left(\frac{-c \kappa ^2+2 c \rho
   ^2+c+2 \Delta  \rho }{2 c \rho +2 \Delta }\right),\\\nn
&&\ell_+=\frac{z_h}{2}  \log \left(\frac{c^2 \gamma ^4-4 \Delta 
   \left(c s \left(\kappa ^2-2 \rho ^2+1\right)+\Delta
   +\Delta  \left(\rho ^2-2\right) s^2\right)}{c^2 \gamma
   ^4-4 \Delta ^2 (\rho  s-1)^2}\right),\\
   &&\ell_-=\frac{z_h}{2 \kappa } \log
   \left(\frac{(c \kappa +\Delta  s)^2}{\rho ^2 s^2-\kappa
   ^2}\right).\eea

It is useful to consider  the difference $\Delta S$ between the entanglement at the current  time moment $S(\ell,\tau)$ and the final state entanglement entropy value $S_{eq}$
 \be
\Delta S(\ell,\tau)=S(\ell,\tau)-S_{eq}(\ell),
\ee
where $S_{eq}$ is defined as
\be
 S_{eq}= \log \left(\frac{R}{\varepsilon} \frac{z_h} {\ell}\sinh\frac {\ell} {z_h}\right).
 \ee 
The bulk variables define relative position of the geodesic top $z_*$ and the point where geodesic crosses the null shell $z_c$ with restrictions  $z_*<z_H$ and $z_c<z_*$.
There are still 3 types of  ETEBA geodesics as in \cite{Balasubramanian:2011ur,HL,HLlong}.

$\,$

For $\tau<0$ our  ETEBA geodesic (first type) lies entirely in the BTZ bulk with temperature $T_i$. The entanglement entropy in this case
is independent of $\tau$ and equals to
\be 
S_i= \log \left(\frac{R}{\varepsilon} \frac{z_H} {\ell}\sinh\frac {\ell} {z_H}\right),
\ee
where $\varepsilon$ is the UV regularization.  
  For $\tau<0$ the limit of large interval length $\ell\rightarrow 
 \infty$ corresponds  to $z_* \rightarrow z_H$ (here
 we mean that $z_* \rightarrow z_H-0$ ) in accordance with \cite{Hubeny:2013hz}.  Let us fix $\ell$ and start to increase the time $\tau$. 

At very small $\tau>0$, the ETEBA geodesic starts intersecting the null shell and for $\tau\ll z_h$
the point of intersection is close to the boundary, $z_c\ll z_h$. This is the second type of the ETEBA geodesics. 

When $\tau$ is of order  $z_h$ the ETEBA geodesic (third type) intersects  the shell behind the horizon, i.e. $z_c>z_h$. At some time $\tau=\tau_s$ the ETEBA geodesic lies entirely in the black hole (with the temperature $T_f$) region.
The role of the second horizon located at $z=z_H$ is that it  pulls out  of the ETEBA geodesic with the top $z_*>z_h$  to the first horizon,
 and this effect is stronger as the difference of two temperatures  decrease, i.e. $\kappa\to 1$.

 The described ETEBA geodesics length  also is related to the equal-time two-point correlation function. The equal-time two-point function $\langle{\cal O}(x,t){\cal O}(x^{\prime},t)\rangle$ of operators ${\cal O}$ with large conformal dimension $\Delta$ between points on the boundary $x$ and $x^{\prime}$ at a time moment $t$ is expressed as 
\bea 
G(x,x^{\prime},t)=e^{-\Delta {\cal L}(x,x^{\prime},t)},
\eea 
where ${\cal L}(x,x^{\prime},t)$ is the corresponding geodesic length. The geodesics described above are responsible for different regimes of behavior of the two-point correlator.


  \subsection{Critical curve}\label{Sect:critline}
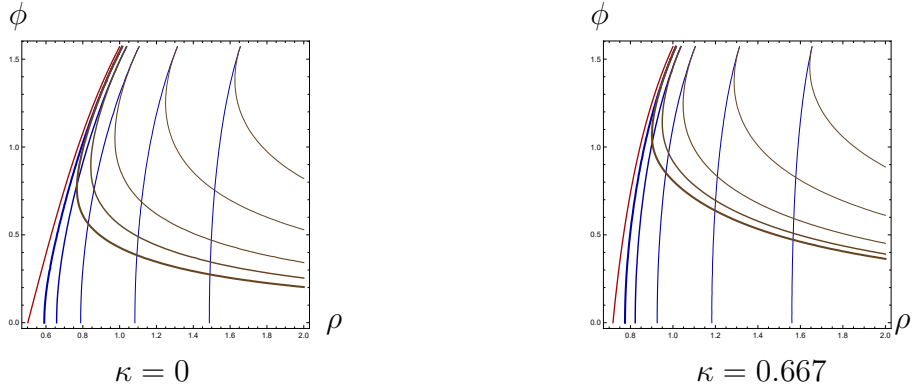
\begin{figure}[h!]
\centering
\begin{picture}(480,130)
\put(80,0){\includegraphics[width=4cm]{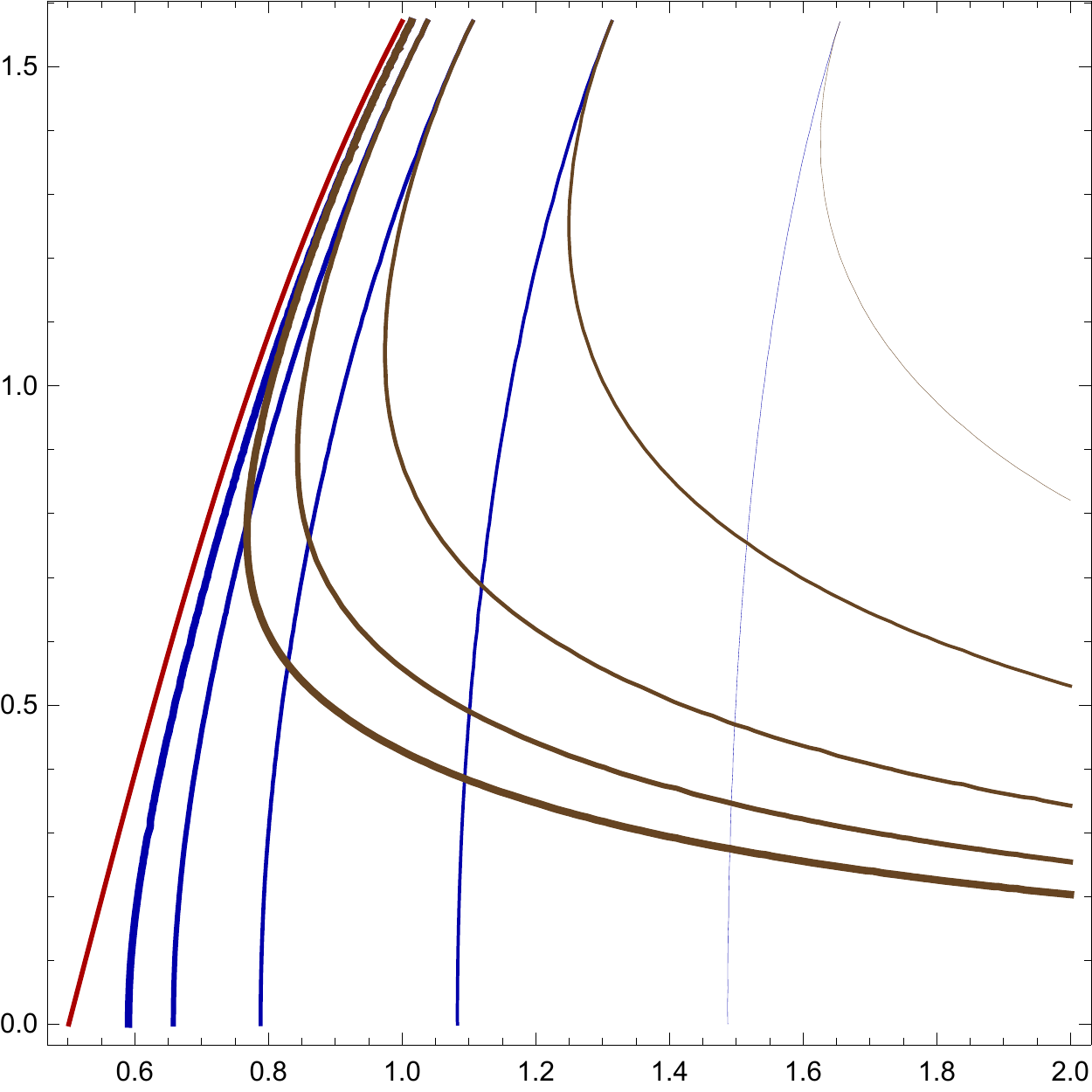}}

\put(300,0){\includegraphics[width=4cm]{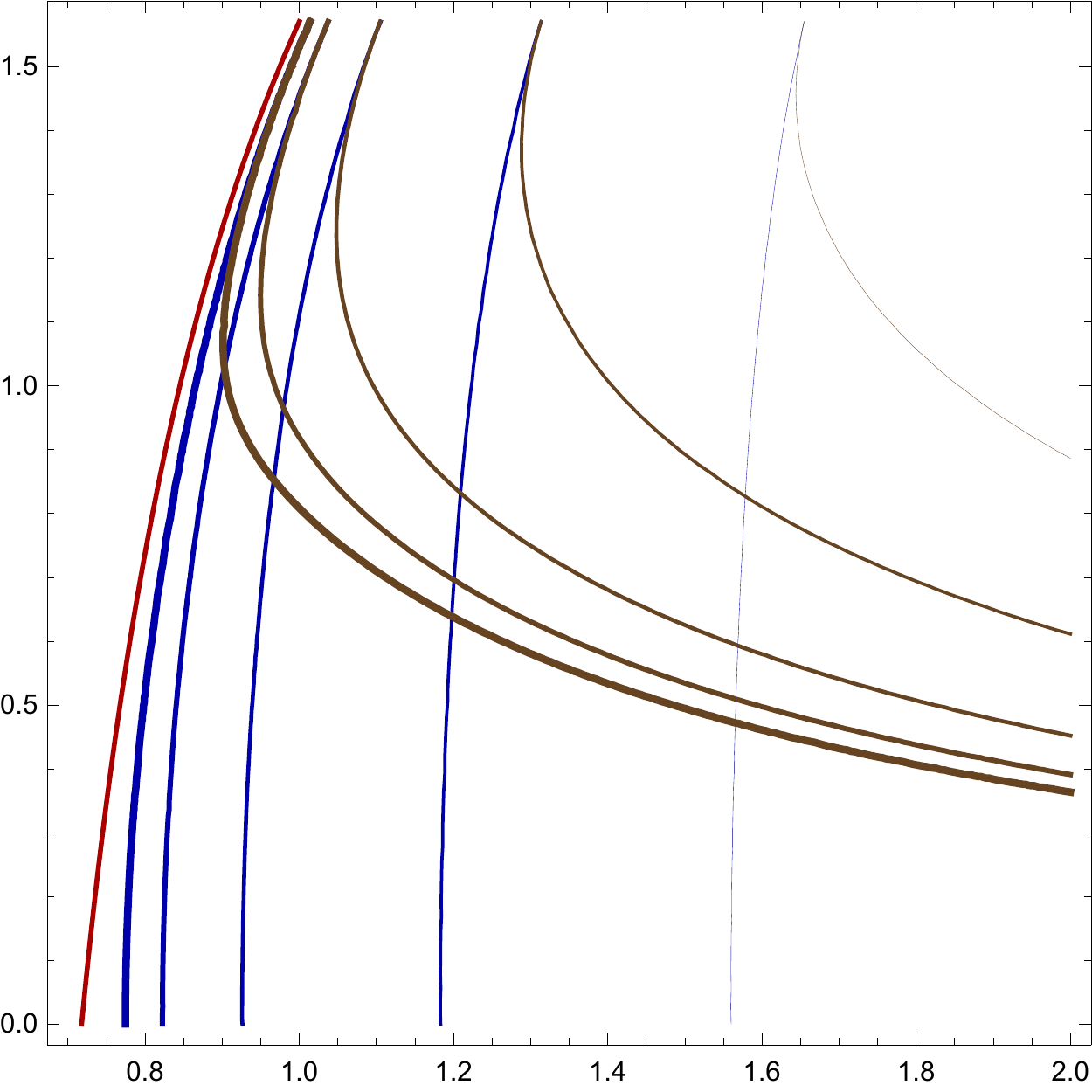}}
\put(300,120){$\phi$}
\put(80,120){$\phi$}
\put(415,5){$\rho$}
\put(200,5){$\rho$}
\put(120,-15){$\kappa=0$}
\put(340,-15){$\kappa=0.667$}
\end{picture}\\$\,$\\
 \caption{
 The contour plots  for $\ell$  and $\tau$ as functions of $\rho$ and $\phi$ for different $\kappa=0$ and $\kappa=0.667$. These plots show that on the right of the critical line (the darker red line) for $\ell>\tau$
 we can always find solution to the system of equations: $\tau=\tau(\rho,\phi); \ell=\ell(\rho,\phi)$.  Lines of the equal times $\tau=0.7,\,1,\,1.5,\,2,\,2.5$ in all panels are shown by blue color with increasing thickness  from  right to left.  Lines of the equal  $\ell=0.7,\,1,\,1.5,\,2,\,2.5$ in all panels are shown by brown  with increasing thickness  from  up to bottom.}
 \label{fig:Change-var}
\end{figure}

  Formula  \eqref{t} gives the explicit dependence for $\ell$ and $\tau$ from the geometry of the entanglement surface specified by parameters $\rho$ and $\phi$.
  For any $0\leq\kappa\leq 1$ there is a critical curve in the parameter space $\rho$ and $\phi$, so that only on the right of this  line (the red line in Fig.\ref{fig:Change-var}) one can perform the single-valued change of variables $(\rho,\phi)\to (\ell,\tau$). To visualize this change of variables  it is convenient to draw the lines of fixed values of $\ell$ and $\tau$ (brown and blue lines in Fig.\ref{fig:Change-var}). From Fig.\ref{fig:Change-var} we see that  large values of $\ell$ and $\tau$ are located near the critical line. The explicit form of the critical line $\rho=\rho_*(\phi)$ is defined by the function $\rho_*=\frac12 {\cal V}(\kappa, \phi)$,
   \bea\label{Vr}
{\cal V}(\kappa, \phi)\equiv1+(1-{\cal Q}) \csc\phi, \,\,\,\,\,\,
{\cal Q}(\kappa, \phi)\equiv\sqrt{\left(1-2 \kappa ^2\right) \cos ^2\phi+2 \kappa ^2 (1-\sin \phi)}.
  \eea
Substituting $\rho_*$ into \eqref{t}  one can check, that  $\tau$ and $\ell_+$ are equal to infinity on the critical line.
Near the critical line we have
\bea\nn&\,&\frac{-c \kappa ^2+2 c \rho ^2+c+2 \rho  \sqrt{\rho
   ^2-\kappa ^2}}{2 c \rho +2 \sqrt{\rho ^2-\kappa ^2}}\Big|_{\rho= \rho^*+\epsilon}=1+2\epsilon {\cal K}_\tau(\phi,\kappa)+{\cal O}(\epsilon^2 ),
      \eea
where
\bea
&&{\cal K}(\kappa,\phi)=\frac{c^2 \Delta  \left(\kappa ^2+2 \rho ^2-1\right)+c \rho  \left(-3
   \kappa ^2+4 \rho ^2-1\right)+2 \Delta ^3}{4 \Delta  (c \rho +\Delta
   )^2}\Big|_{\rho=\rho_*}\nn,\eea
   and, therefore,  we   get  the following time asymptotic near the critical curve \bea\label{t-near-critl}
&& \tau = -\frac{z_h}{2} \log \varepsilon-\frac{z_h}{2} \log {\cal K}+{\cal O}(\varepsilon).
\eea
The  expansion for $\ell_+$ takes the similar form
\be
 \ell_{+}\approx-\frac{z_h}{2}\log \epsilon+\frac{z_h}{2}\log {\cal K}^+,
 \ee
  where the function ${\cal K}^+=\frac{\fN}{\fK}$ is defined as
 \bea
\fK&=&2\sqrt{2}\,{\cal Q}\,\sin \phi -4\cos^2\phi +4\kappa^2(2\sin\phi(1-\sin \phi )-{\cal Q}),\nn\\\nn
\label{KL+s1}
\fN
&=& 4\left(\kappa ^2-1\right)^2 \cos ^2\phi
   -\frac{\left({\cal V}^2-4 \kappa ^2\right) \left(\left({\cal V}^2-8\right) \sin ^2\phi
   +4\right)}{4}
   \label{KLsmooth}
   +\\&+&\frac12 \sqrt{{\cal V}^2-4 \kappa ^2} \left({\cal V}^2-2 \left(\kappa
   ^2+1\right)\right) \sin (2 \phi ),
\eea
where ${\cal V}$  and  ${\cal Q}$ is defined by \eqref{Vr}.

The expression for $\ell_{-}$ is not singular on the critical line
\be
\label{ellm}
\ell_{-}=\frac{z_h}{2 \kappa}\log \cal K^{-},\,\,\,\,\,\,\,\,{\cal K}^{-}=
   \frac{\text{2} \sin \phi  \sqrt{ {\cal V}^{\text{2}}-\text{4}\kappa^{\text{2}}}+\text{4}\kappa\cos\phi}{{\cal V}^{\text{2}} \sin^{\text{2}} \phi -\text{4}\kappa^\text{2} }.\ee

   In the limit of $\kappa\to 0$ we reproduce the expressions from  \cite{HL}
   \bea
    {\cal K}(\kappa,\phi)\underset{\kappa\to 0}\to 1 +\cot \frac{\phi}{2},\,\,\,\,\,\,\,\,\,\,
    {\cal K}^+(\kappa,\phi)\underset{\kappa\to 0}\to
    \frac{\cot \frac{\phi}{2}+1}{\cot \frac{\phi}{2}},\,\,\,\,\,\,\,\,\,\,
   {\cal K}^-(\kappa,\phi)\underset{\kappa\sim 0}\approx 1.\eea
   
      \section{Universality of critical behaviour}\label{Sect:MLR}
   \subsection{Regimes in holographic heating}\label{Sect:MLR1}

\begin{figure}[h!]
\centering
\includegraphics[width=12cm]{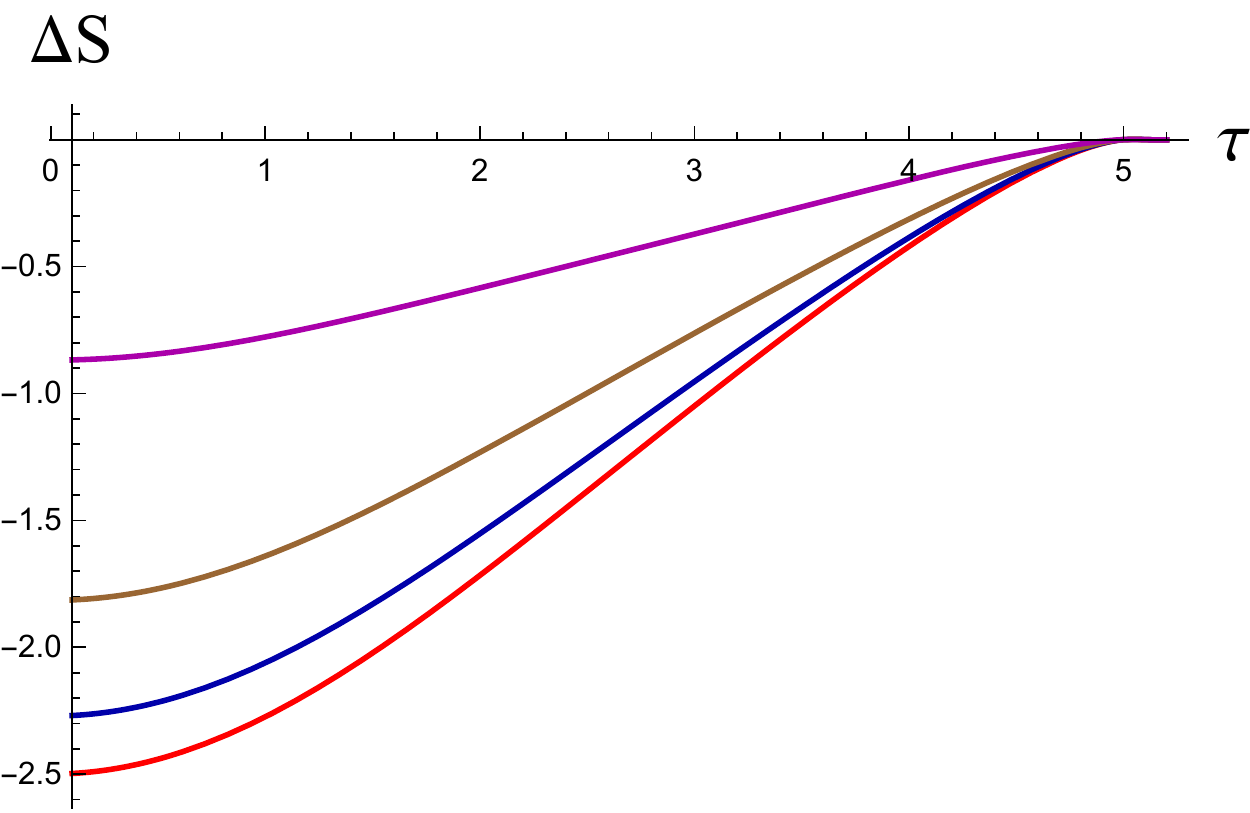}
 \caption{The dependence of  $\Delta S$ on time $\tau$ for $\ell=5$ for different values of $z_H=\infty,3,2,1.3$ from down to top.}
 \label{fig:F}
\end{figure}

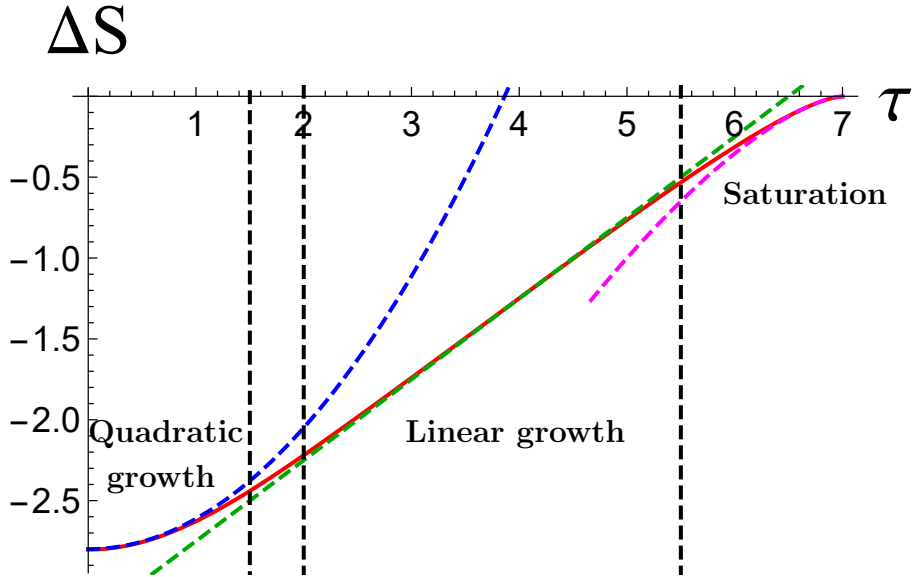
\begin{figure}[h!]
\centering
\begin{picture}(0,0)
\end{picture}\\$\,$
\put(-200,0){\includegraphics[width=12cm]{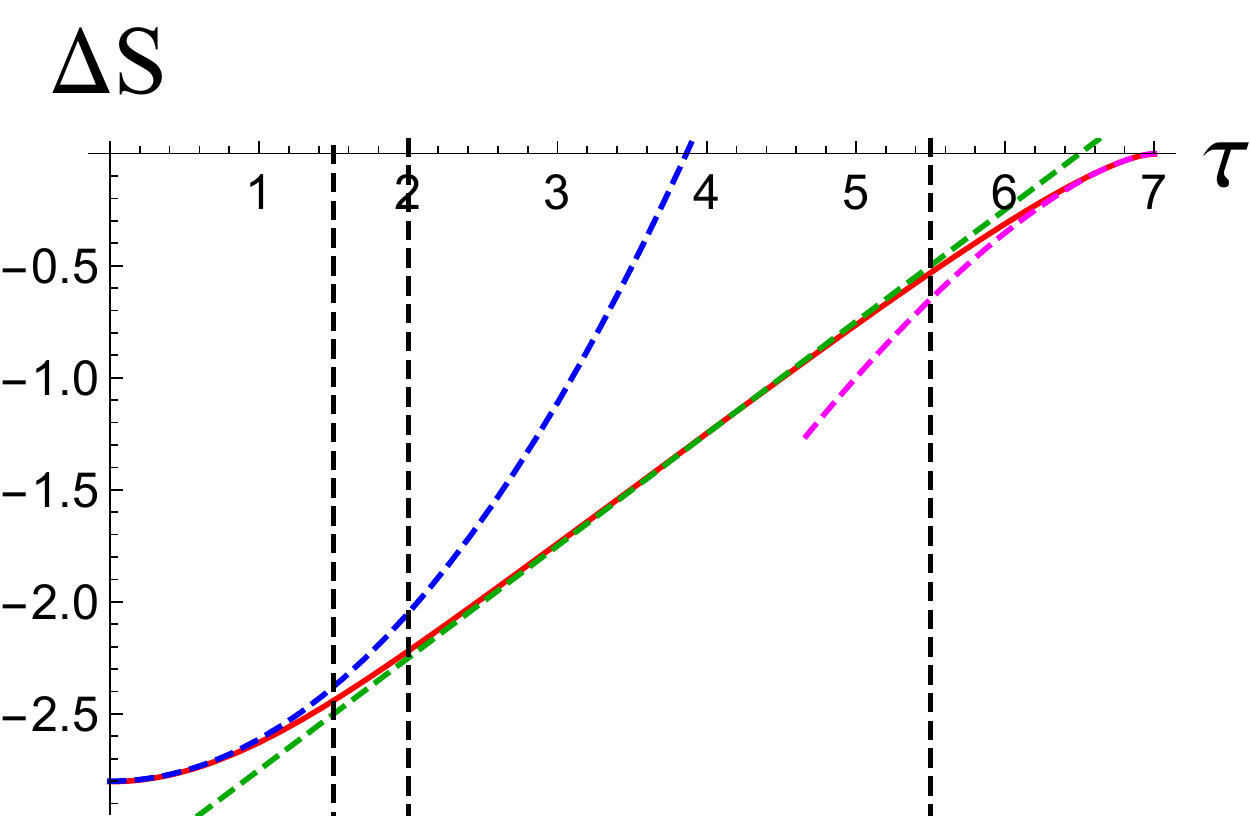}}
\put(-170,50){\textbf{Quadratic}}
\put(-163,33){\textbf{growth}}
\put(-50,50){\textbf{Linear growth}}
\put(70,140){\textbf{Saturation}}
 \caption{Typical time dependence of $\Delta S(\ell,\tau)$ for some fixed $\ell$ and $z_H$ (here $\ell=7$ and $z_H=2$). The red curve is the entanglement entropy dependence $\Delta S(\ell,\tau)$, the blue curve is the quadratic approximation to initial growth \eqref{quadDS}, the green line is the linear growth regime \eqref{spread} and magenta line corresponds to the asymptotic describing saturation regime \eqref{appsat}.}
 \label{fig:regimes}
\end{figure}

In this subsection we establish different regimes of heating using the asymptotic behavior of our explicit formula \eqref{S-AA}. The regimes classification and notion of memory loss for BTZ-Vaidya model have been considered in \cite{HL}. The equilibration starting from initial thermal state shares all basic regimes of thermalization process.  There are the following regimes  corresponding to different asymptotics for $\Delta S(\ell,\tau)=S-S_{eq}$. For time dependence of the entanglement entropy corresponding to different initial temperatures see Fig.\ref{fig:F}. Also see Fig.\ref{fig:regimes}  below where we plot the comparison for  approximations of different regimes asymptotics on timescale versus exact formula.  It is worth to note a   smooth character of the saturation regime (magenta and red lines in Fig.\ref{fig:regimes})

 Namely
 
  \bi \item {\it Pre-local-equilibration growth} regime is considered in \ref{quadrgrowth}
 
  \item {\it Memory loss} and subregimes of this regime is the main subject of Sect.\ref{MLR3}. Post-local-equilibration linear growth is considered in  \ref{MLR:lin}, the saturation subregime is the subject of subsection \ref{MLR:S} and  late-time memory loss is briefly discussed in \ref{MLR:late}.
 \ei 

\subsection{Pre-local equilibration growth}\label{quadrgrowth}
In \cite{HL} it was found, that the entanglement entropy in the system with zero initial temperature following the sharp quench first of all exhibits the period of quadratic growth. As in zero initial temperature case the quadratic growth regime is also present for $T_i>0$. This regime takes place before the local equilibration defined by the scale $\xi\approx 1/T_f$ is established. 
   Pre-local equilibration growth regime occurs for small $\tau$, namely when $\tau \ll z_h$.  Here we consider  this regime assuming that $\ell \rightarrow 0$.

Using equations \eqref{t} when $\phi \to 0$ and $\tau/z_h \to 0$ we get the asymptotic for $\rho$ corresponding to this limit

 \be
    \label{rhoB}\rho\approx\frac{z_h}{\tau}+B\,\frac{\tau}{z_h},\,\,\,\,\,\,B=\frac{ 3 \kappa ^2 +1}{12}.
    \ee
Also from \eqref{t} we get the expansion for $\ell$ in the form
 \bea\label{ell-t}
\ell \approx \frac{t}{\sin\phi}+  \frac{\kappa^2t^3}{3z_h^2\sin^3\phi} -\frac{3\kappa ^2 +1}{12}  \frac{t^3}{z_h^2\sin\phi} .
 \eea
 
 From equation \eqref{S-AA}, and expansions \eqref{rhoB} and \eqref{ell-t} we get that $\Delta S$ (in the limit $\ell \rightarrow 0$) has the form
 \bea\label{quadDS}
&&\Delta S \approx\fg_\kappa \cdot \tau^2,\\
&&\fg_\kappa=\frac{\left(1-\kappa^2\right)}{4 z_h^2} .
\eea
This is the behavior of entanglement entropy in pre-local equilibration growth regime also known as quadratic growth regime.  We explicitly observe the dependence on initial temperature. One can rewrite \eqref{quadDS} using \eqref{temperatures}  in the form
\be
\Delta S \approx 2 \pi \left({\cal E}_f-{\cal E}_i \right)\,\tau^2.
\ee
  Similar to the thermalization case before the local equilibration of the system the data characterizing
the initial growth is the energy density difference between initial and final state.
\subsection{Memory loss regime}\label{MLR3}
  {\it Memory loss} and subregimes of this regime \cite{HL} are the main subjects of this section. This regime occurs when both $\tau$ and $\ell$ are larger than $z_h$
  \be\nn
z_h\ll \tau\leq\ell.\ee
Memory loss means that after the time when local equilibrium is established the entanglement propagation has the dependence only on difference $\ell-\tau$ instead of $\ell$ and $\tau$ separately.
 Depending on the  ratio of $\ell-t$ to $\ell$, $\tau$, $z_h$ there are the following more detailed subregimes cases.
The first subregime we discuss is {\it post-local-equilibration linear growth}, with entropy dependence
\be
\Delta S\approx(1 -\kappa) (\ell-\tau),
\ee 
  where $\ell$ is large and $z_h\ll  \tau\ll\ell$.
Then we consider  the {\it saturation } regime, characterized by the entanglement entropy dependence 
\be
\Delta S\approx  -\ff_\kappa\,\left(\frac{\ell-\tau}{z_h}\right) ^{3/2}-\fn_\kappa\, \left(\frac{\ell-\tau}{z_h}\right) ^2,
\ee
 where coefficients $\ff_\kappa$ and $\fn_\kappa$ depend on the initial temperature. These coefficients are calculated in Sect.\ref{MLR:S}. The validity of this regime is provided when $\ell-\tau\ll z_h.$
The last regime is the  {\it  late-time memory loss } regime. It occurs when $
\ell\gg \ell -\tau\gg z_h.$
We briefly discuss this regime in Sect.\ref{MLR:late}.

 Using formulae \eqref{t-near-critl} and \eqref{ellm}  we get, that near the critical curve the difference 
$T_-\equiv 1/z_h\left(\tau-\ell\right)$ is finite and can be expressed in the form
   \bea\nn\label{Tm}
T_-&=&- \chi_{\kappa}(\phi),
\eea
where function $\chi_\kappa$  is defined as
\bea\nn
\chi_{\kappa}(\phi)\equiv-\frac12\log{\cal K}(\kappa,\phi) {\cal K}^+(\kappa,\phi) - \frac{1}{2\kappa}  \log {\cal K}^{-}(\kappa,\phi).\eea
 The second light cone variable $T_+\equiv  \tau/z_h +\ell/z_h$ is singular near the critical line $T_+\sim -\log \epsilon$.
The function $\chi_{\kappa}(\phi)$ plays an important role in what follows, controlling the behavior of the light cone variable 
$T_-$. In the limit $\kappa\to 0$
we have 
$\chi_{\kappa}(\phi)\underset{\kappa\to 0}{\to}\chi_{0}(\phi)$,
 where
\be 
 \chi_{0}(\phi)=\cot\frac{\phi}{2}-1 - \log \cot\frac{\phi}{2}
 \ee
 is in accordance with \cite{HL}. The inverse function $\chi^{-1}$ can be expressed in terms of Lambert function $W$
\be
\phi=\chi_{0}^{-1}(T_-)=2{\mbox{arccot}}\left(-W_{-1}\left(-e^{T_--1}\right)\right).\ee
Now let us consider the entanglement entropy behavior near the critical line
\bea\label{s-s1}
&&S-S_{eq}=   \log \frac{1}{\fS}\frac{\sinh\frac{\tau}{z_h }}{\sinh\frac{\ell}{z_h }}
\approx \frac{\tau-\ell}{z_h }-2\sinh  \frac{\tau-\ell}{z_h }e^{ -\frac{\tau+\ell}{z_h }} 
   -\log \fS(\phi,\rho^*_H,\kappa).\nn
\eea
Note, that $\fS(\phi,\rho,\kappa)$ is not singular on the critical line.  Neglecting the exponentially small terms in \eqref{s-s1} and introducing $\fS_{\kappa}(\phi)={\fS}(\phi,\rho^*_H,\kappa)$ we get
\bea
\Delta S\equiv S-S_{eq}&\approx&T_-
   -\log {\fS}_{\kappa}(\phi).
   \eea
   Formula \eqref{Tm} gives the representation of $\phi$ in terms of $T_-$ as 
  $\phi=\chi^{-1}_\kappa(-T_-)$.
  Near the critical line the entanglement entropy has the wave spreading
  \bea\label{spread}
\Delta S&\approx& T_-
   -\log \fS_{\kappa}(\chi^{-1}_\kappa(-T_-)).
   \eea
The dependence of entanglement entropy propagation only on  $T_-$, i.e. realization of the memory loss regime, 
is based on the exponential suppression of $T_+$ dependence, see \eqref{s-s1}, and dependence of $\fS$
only on $\phi$ near the critical line.
 

\subsubsection{Post-local-equilibration linear growth.  Expansion near $ \vartheta_\kappa$}\label{MLR:lin}
Above we established that the memory loss regime corresponds to the specific behavior of $\Delta S$, $\ell$ and $\tau$ in parametric space $\rho$ and $\phi$ near the critical curve resulting in dependence only on difference $\ell-\tau$. To pick out the specific behavior of entanglement corresponding to linear growth subregime let us analyse the behavior of $\ell$ on the critical curve.
First let us consider $\ell_-$.
The new feature  of behavior of $\ell_-$ on the critical curve for $\kappa>0$,
 is that it is singular near  $\phi=\vartheta_\kappa$, where 
 \be \vartheta_\kappa=\arccos\left(\frac{\sqrt{1+2 \kappa -3 \kappa
   ^2}}{\sqrt{1+2 \kappa +\kappa ^2}}\right).
   \ee
When $\phi= \vartheta_\kappa+\delta$ we get
\bea
\frac{\ell_-}{z_h}\approx -\frac{ \log
   \delta }{2 \kappa }+\zeta^-_\kappa,\,\,\,\,\,\,\,\,\, \zeta^-_\kappa=
\frac{ 1}{2 \kappa }\log \left(\frac{4 \kappa  \left(-2 \kappa
   ^2+\kappa +1\right)}{(\kappa +1)^2 \sqrt{(2-3 \kappa )
   \kappa +1}}\right).
\eea

Since  time $\tau$ and $\ell_+$ are not singular near this point we get
\bea
\label{chi-appr}
&&-T_-= \chi_{\kappa}(\vartheta_\kappa+\delta)
=-\frac{ 1}{2 \kappa }\log
   \delta 
+\zeta_\kappa,\,\,\,\,\,\,\,\,\zeta_\kappa=\zeta_\kappa^-+\frac12\log\frac{\kappa^2}{(1 + \kappa)^2}.\eea
The function  $\fS$ from equation \eqref{fS} has the expansion near the critical line and at $\phi\to\vartheta_{\kappa}+\delta$,
 \be
  \fS_\kappa (\rho,s)\Big|_{\rho=\rho_{cr},\phi=\vartheta_\kappa+\delta}= \frac{2\fq_\kappa} {\fr_\kappa}\delta^{1/2},\ee
where $\fq_\kappa$  depends on $\delta$ and $\fr_\kappa$ is some constant. We do not need to determine them explicitly. For completeness as an example at $\kappa=0.25$ we have  $\fq_{0.25 }=0.39 \delta$ and $\fr_\kappa$ is approximated by $ \fr_\kappa =1.331\, -3 (\kappa -0.334)^2$.

  Hence we have the expansion of $\Delta S$ 
   \bea\label{spread}
\Delta S&\approx& 
   T_-
   -\log \frac{2\fq_\kappa} {\fr_\kappa}e^{-\kappa  (-T_--\zeta_\kappa )})=
  (1 -\kappa) T_-
  -\kappa \zeta_\kappa  -\log \frac{2\fq_\kappa} {\fr_\kappa}.
   \eea 
   Therefore, the linear coefficient is changed as compare to the case of the initial zero temperature, and  it is
   \be
   \label{fn}
   \fk_\kappa=1-\kappa.\ee
The coefficient $\fk_\kappa$ is identified with the entanglement velocity \cite{HL,Mezei1}. The coefficient $\fk_0=1$ corresponds to $T_i=0$ and this is one of the fundamental facts about two-dimensional CFT. The fact, that initial temperature decreases this coefficient, can have two different explanations, see \cite{Mezei1}. The first one is the presence of quasiparticles, streaming from the quench start with the speed less than CFT speed of sound.  The second one is the consequence of particle multiple re-scattering and interactions. See \cite{Cardy:2015xaa,Verlinde} for description of CFT in  thermal state deformation by interaction. Now let us define $W(t)$ as a state of the half-space of our system at  moment $t$. Then we define $w(t)$ as a  boundary point of this half space. Finally define $X$ to be the null vicinity of point $w(t)$. The coefficient $\eqref{fn}$ also give rise to connection (see formula 2.9 in \cite{Mezei1} and general discussion) with  the mutual information $I(W(t),X)$ 

\bea\label{mut}
\kappa \sim I(W(t),X).
\eea

 This indicates presence of entanglement inside the effective emergent light cone between EPR pairs produced by quench.


\subsubsection{Saturation. Expansion near $\pi/2$}\label{MLR:S}

The equilibration process ends with the saturation regime. In \cite{HLlong,Hubeny:2013hz} the dependence of entanglement entropy in this regime for 2d holographic CFT for  $T_i=0$ was found. We generalize this result on the case when initial temperature is not zero. Here we naturally consider the saturation as one of the limiting forms of memory loss regime, namely limit  $T_-\to -0$ in  \eqref{spread}. This corresponds to time scales near the final equilibration time $t_{s}\approx \ell$. In parametric space this regime corresponds to the values $\phi\to  \pi/2$. 

Let us consider asymptotics of $\tau$ and  $\ell=\ell_-+\ell_+$  given by \eqref{ellm} and \eqref{t-near-critl} when $\phi \to \pi/2$. In this limit the asymptotic expansion for the combination $\ell-\tau$ takes the form
\bea
\label{appr-pi2}
&&\frac{\ell-\tau}{z_h}= \chi_{\kappa}(\phi)\approx  \frac{z_h \delta ^2}{2 \left(1-\kappa
   ^2\right)}+\frac{\left(z_h+2 \kappa ^2 z_h\right) \delta
   ^3}{6 \left(1-\kappa ^2\right)^{3/2}}-\frac{\left(5 z_h+7
   \kappa ^2 z_h\right) \delta ^4}{24 (1-\kappa )^2 (1+\kappa
   )^2},
\eea
where $\delta \approx \pi/2-\phi$.
 The inverse function $\chi^{-1}$ is expressed in the form
  \bea\nn
 \delta\approx-\sqrt{2(1-\kappa ^2)}\Big(\frac{\ell-\tau}{z_h}\Big)^{1/2}&+&
 \frac{\sqrt{1-\kappa^2}(1+2\kappa^2)}{3}\frac{\ell-\tau}{z_h}
 -\\&&-\frac{ \left(20 \kappa ^4-\kappa
   ^2-10\right) \sqrt{1-\kappa ^2}}{18 \sqrt{2}}\left(\frac{\ell-\tau}{z_h}\right) ^{3/2}.
 \eea
Finally,  expanding  $\log \fS$ near $\phi \to \pi/2$
\bea
&&\log \fS \approx \frac{\delta^2}{2 \left(1-\kappa
   ^2\right)}+\frac{\kappa ^2 \delta^3}{2
   \left(1-\kappa ^2\right)^{3/2}}+\frac{\left(2+7 \kappa ^2+3 \kappa ^4\right) \delta^4}{24 \left(1-\kappa ^2\right)^2},
\eea
we get
\bea\label{appsat}
\Delta S&\approx & -\ff_\kappa\,\left(\frac{\ell-\tau}{z_h}\right) ^{3/2}-\fn_\kappa\, \left(\frac{\ell-\tau}{z_h}\right) ^2\\\nn
\ff_\kappa&=&  \frac{\left(\sqrt{2} \left(1-\kappa^2\right)\right)
 }{3 },\,\,\, \fn_\kappa= \frac{(1-\kappa^2)^2}{6 }, \eea
which generalizes the result obtained in \cite{Hubeny:2013hz,HL}.  
   We see that there is essential dependence of  the coefficient $\ff_\kappa$ in front of scaling law  on the initial temperature.
  Note, that at  $\kappa=1$ all the expansion for saturation regime vanishes identically. The initial temperature dependence of the saturation regime involves only powers of $1-\kappa^2$.


 \subsubsection{Late-time memory loss regime.  Interpolation between $\phi=\vartheta_\kappa$ and $\pi/2$}\label{MLR:late}
There is
an additional regime  taking place on the timescale between linear growth
and saturation interpolating between them. It is called \cite{HL} "late-time memory loss".
In this regime the entanglement entropy depends only on time remaining till saturation.  We outline some approximate formulae to describe this regime.
As   interpolation functions $\fx_\kappa(\phi)$ to $\chi_\kappa(\phi)$ on the interval $(\vartheta_\kappa,\pi/2)$
we take
\bea\nn
\fx_\kappa&=&a_{\kappa}\Big(\cot \frac{\phi-\vartheta _\kappa}{2b_\kappa}
-\cot \frac{\frac{\pi}{2}-\vartheta_ \kappa}{2b_\kappa}
-\log \frac{\cot \frac{\phi-\vartheta _\kappa}{2b_\kappa}}{
\cot \frac{\frac{\pi}{2}-\vartheta _\kappa}{2b_\kappa}}\Big),
\label{interp}
\eea
where the numerical values of $a_\kappa$ and $b_\kappa$ are some numerical constants, for example $a_{0.1}=0.15$ and $b_{0.1}=2.5$.
The inverse functions to $\fx_\kappa$ are given by
\bea
\label{inver-chi}
&&\fx_\kappa^{-1}\equiv\phi=\vartheta_\kappa+2b_\kappa{\mbox {arccot}} W_{-1}\Big(-e^{-\frac{\fx_\kappa}{a_\kappa}-\cot \phi_{\kappa,0}}\cot \phi_{\kappa,0}\Big),\nn\eea
where we define $\phi_{\kappa,0}$ as
\be
\phi_{\kappa,0}\equiv \frac{\frac{\pi}{2}-\vartheta_\kappa}{2b_\kappa}.
\ee
Substituting $\phi_{\kappa,0}$ in \eqref{spread} we get
  \bea\label{spread-f}
&&\Delta S\approx T_-
   -\log \fS_{\kappa}\left(\vartheta_\kappa+2b_\kappa{\mbox {arccot}} W_{-1}\Big(-e^{\frac{T}{a_\kappa}-\cot \phi_{\kappa,0}}\cot \phi_{\kappa,0}\Big)\right).
   \eea


\newpage
\section{Conclusions and discussions}\label{Sect:Final}

In this paper using holographic approach, we have considered the evolution of entanglement entropy of single interval
during equilibration after the global sharp quench of the initial thermal state at temperature $T_i$. This quench is followed by non-equilibrium heating up of the system to temperature $T_f$. We use Vaidya thin shell in the BTZ black hole background (so-called double-BTZ-Vaidya)  as a holographic model of the process. The main purpose of this paper is to compare global quench process starting from  thermal initial state with the one starting from the vacuum state.
We have shown that the quench starting from thermal initial state shares all qualitative features of zero initial temperature case, but there is a quantitative dependence on $\kappa=T_i/T_f$. 

We have calculated the critical exponents corresponding to  various regimes of entanglement entropy propagation. These exponents  turn out to be independent on temperature. However, the coefficients $\ff_\kappa$, $\fn_\kappa$,  $\fk_\kappa$ and $\fg_\kappa$ (see  \eqref{quadDS}, \eqref{fn} and  \eqref{appsat}) corresponding to these exponents are $\kappa$-dependent.
Namely,
in section \ref{quadrgrowth} we have derived the corrections to quadratic growth regime that occurs before local equilibrium is set.    The pre-local-equilibration stage in non-equilibrium  heating  is very similar to one during thermalization \cite{HLlong}.
 In our case the early growth time dependence is proportional to the difference of the energy densities
\be
\Delta S=2 \pi \left({\cal E}_f-{\cal E}_i \right)\,\tau^2 +...,\ee
that is consistent with   the initial evolution of  HEE  with $T_i=0$ \cite{Allahbakhshi:2013rda,Bhattacharya:2012mi,Blanco:2013joa,Wong:2013gua,HL,HLlong}. For small interval size, the local equilibration scale $\xi\approx 1/T_f$ is independent on the  initial temperature.

  In section \ref{MLR3}, using the explicit formula for the evolution of the HEE, we have shown the existence of the   memory loss regime in  non-equilibrium heating.
 The HEE evolution in this regime  is described by the function of one variable 
   \bea\label{spread-m}
\Delta S(\ell,\tau)&\approx&\fM_\kappa(\ell- \tau),
   \eea 
   with an explicit dependence of the coefficient $\fM_\kappa$ on $\kappa$.
 The memory loss regime occurs long after the system has achieved local equilibration at scales of order $z_h$. However, we have found that  details of memory loss explicitly depend on the initial temperature. Thus we find that only geometric 
data  are lost when the system follows this regime, while some initial state details, like temperature, are resistant  to be erased. 

 Similar to the thermalization \cite{HL}, there are two special cases of the memory loss regime. 
In Sect.\ref{MLR:lin} we have derived corrections to {\it post-local-equilibration linear growth} 
 with large $\ell$, i.e.  $z_h\ll  \tau\ll\ell$. In this regime
\be\label{PLELG}
\fM_\kappa(\ell-  \tau)\approx -\fk_\kappa(\ell-  \tau)+...,\ee
 the scaling parameter $\fk_\kappa$ depends on $\kappa$ as 
\be \fk_\kappa=1-\kappa.
\ee 
This coefficient can be identified with the entanglement tsunami velocity. Following \cite{Mezei1} there are two possible explanations for 
the reduction of this velocity. The first one may be related with the presence of quasiparticles, which travel with the speed less than the effective speed of the light. The second one, is due to multiple interactions between quasiparticles and interactions with thermal fluctuations.  It is possible to interpret the decrease in the speed of a tsunami by the propagation of entanglement inside the light cone. The value on which this speed decreases allows an interpretation in terms of the mutual information.

Section \ref{MLR:S} is devoted to calculation of the corrections to critical exponents at the saturation regime. This regime takes place when
  $\ell-\tau\ll z_h$, and
\be
\fM_\kappa(\ell- \tau )\approx  -\ff_\kappa\,\left(\frac{\ell-\tau}{z_h}\right) ^{3/2}-\fn_\kappa\, \left(\frac{\ell-\tau}{z_h}\right) ^2+...\ee
where the scaling parameter $\ff_\kappa$  depends on $\kappa$ as 
\be \ff_\kappa=  \frac{\sqrt{2}}{3} \left(1- \kappa ^2\right)
\ee and 
\be\fn_\kappa= \frac{1}{6}(1-\kappa^2)^2. 
\ee

To summarize, we have shown that entanglement entropy  evolution after the quench with the thermal initial state retained all regimes of the entanglement entropy evolution after the quench with vacuum initial state. All  these evolution regimes, except saturation,  are   also present in the quench with a strongly inhomogeneous initial state \cite{AKT} (for holographic description of single local quench see ).

It is worth to note,  that in our model  the equilibration time does not depend on the initial temperature. However,
as we  have seen, the speed of  the entanglement propagation decreases with increasing of the initial temperature.
Hence, one can conclude, that the reason for the entanglement speed reduction is the  interaction of the quasiparticles stream with thermal fluctuations and this interaction is different at different stages of thermalization. In other words, one can say that {\it the system responses to the reduction of the speed of entanglement propagation  preserving full time of equilibration of the  entanglement for the given interval.}

For more complicated models, we  expect that 
  \bea\label{spread-mm}
\Delta S(\ell,\tau)&\approx&\fM_\kappa(\ell-  \fv\tau),
   \eea  
  where the form of  $\fM_\kappa$ depends on the model,
but the dependence  on $(\ell- \fv \tau)$ with different value of $ \fv$,
will survive for more general initial states.
 In particular,  for the linear growth  regime
 we expect  that the speed $\fv$,  which characterizes  properties of the equilibrium state,  is solely determined by  the black hole
 describing the final state. Equation \eqref{spread-mm} manifests itself  the local nature of the  entanglement propagation. 
 s

 It would be interesting to extend the results we have obtained to the HEE and the other nonlocal observables to a higher dimensional cases as well to study the heating  process initiated by more general quenches, in particular  quenches
with inhomogeneous  states (see for \cite{Balasubramanian:2013oga} $T_i=0$ case) or defined by various infalling shells. In particular these shells include
massive infalling shells, charged shells, shells with angular momentum (corresponding thermalization process have been studied in \cite{Danielsson:1999zt, Danielsson:1999fa, Erdmenger:2012xu, 1602.05934, Caceres:2012em, Galante:2012pv}, \cite{ABK}).
Thick shells infalling in the black hole background  in higher dimensional cases have been already used  to study numerically the holographic non-equilibrium heating   \cite{IAIV}. When  the thickness of the shell is less then typical sizes of intervals which we deal with, the evolution of the entanglement entropy for large intervals also  shows the memory loss regime\cite{HLlong}. Also, it would be interesting to compare the results of this work and possible higher-dimensional generalizations with different results concerning equilibration of thermal states in holographic context including the numerical study of black hole with falling scalar field thin shell \cite{Landsteiner:2017lwm} and the holographic description of non-equilibrium  thermal transport in two isolated quantum critical systems with different temperatures is found in \cite{Bhaseen:2013ypa,Erdmenger:2017gdk}.

\section*{Acknowledgements}

The authors are grateful to M. Khramtsov and M.Tikhanovskaya for useful discussions.
This work is supported by the Russian Science Foundation (project 17-71-20154).


\begin{thebibliography}{91}
\bibitem{Malda} 
  J.~M.~Maldacena,
  ``The Large N limit of superconformal field theories and supergravity,''
  Int.\ J.\ Theor.\ Phys.\  {\bf 38}, 1113 (1999)
  [Adv.\ Theor.\ Math.\ Phys.\  {\bf 2}, 231 (1998)]
   [hep-th/9711200].


\bibitem{GKP}
  S.~S.~Gubser, I.~R.~Klebanov, A.~M.~Polyakov,
  ``Gauge theory correlators from noncritical string theory,''
  Phys.\ Lett.\  {\bf B428}, 105-114 (1998) [hep-th/9802109].

 \bibitem{Witten}
  E.~Witten,
  ``Anti-de Sitter space and holography,''
  Adv.\ Theor.\ Math.\ Phys.\  {\bf 2}, 253-291 (1998) [hep-th/9802150].

\bibitem{WittenTH} 
  E.~Witten,
  ``Anti-de Sitter space, thermal phase transition, and confinement in gauge theories,''
  Adv.\ Theor.\ Math.\ Phys.\  {\bf 2}, 505 (1998)
  [hep-th/9803131].

  
  \bibitem{Danielsson:1999fa} 
  U.~H.~Danielsson, E.~Keski-Vakkuri and M.~Kruczenski,
  ``Black hole formation in AdS and thermalization on the boundary,''
  JHEP {\bf 0002}, 039 (2000)
  [hep-th/9912209].

\bibitem{MaldacenaET} 
  J.~M.~Maldacena,
  ``Eternal black holes in anti-de Sitter,''
  JHEP {\bf 0304}, 021 (2003)
  [hep-th/0106112].


\bibitem{CasalderreySolana:2011us} 
  J.~Casalderrey-Solana, H.~Liu, D.~Mateos, K.~Rajagopal and U.~A.~Wiedemann,
  ``Gauge/String Duality, Hot QCD and Heavy Ion Collisions,''
  book:Gauge/String Duality, Hot QCD and Heavy Ion Collisions. Cambridge, UK: Cambridge University Press, 2014
  [arXiv:1101.0618 [hep-th]].
  
  \bibitem{IA}   I.~Ya.~Aref'eva,
  ``Holographic approach to quark-gluon plasma in heavy ion collisions,''
  Phys.\ Usp.\  {\bf 57}, 527 (2014).



\bibitem{DeWolf}
  O.~DeWolfe, S.~S.~Gubser, C.~Rosen and D.~Teaney,
  ``Heavy ions and string theory,''
  Prog.\ Part.\ Nucl.\ Phys.\  {\bf 75}, 86 (2014)
  [arXiv:1304.7794 [hep-th]].



\bibitem{Hartnoll:2016apf} 
  S.~A.~Hartnoll, A.~Lucas and S.~Sachdev,
  `Holographic quantum matter,''
  arXiv:1612.07324 [hep-th].
 
\bibitem{Easther:2011wh} 
  R.~Easther, R.~Flauger, P.~McFadden and K.~Skenderis,
  ``Constraining holographic inflation with WMAP,''
  JCAP {\bf 1109}, 030 (2011)
  [arXiv:1104.2040 [astro-ph.CO]].


\bibitem{AbajoArrastia:2010yt} 
  J.~Abajo-Arrastia, J.~Aparicio and E.~Lopez,
  ``Holographic Evolution of Entanglement Entropy,''
  JHEP {\bf 1011}, 149 (2010)
  [arXiv:1006.4090 [hep-th]].

 \bibitem{Balasubramanian:2011ur} 
  V.~Balasubramanian {\it et al.},
  ``Holographic Thermalization,''
  Phys.\ Rev.\ D {\bf 84}, 026010 (2011)
  [arXiv:1103.2683 [hep-th]].
  
 \bibitem{Lopez}
J.~Aparicio and E.~Lopez,
  ``Evolution of Two-Point Functions from Holography,''
  JHEP {\bf 1112}, 082 (2011)
  [arXiv:1109.3571 [hep-th]].

 
      \bibitem{IAIV} I.Ya.~Arefeva and I.V.~Volovich,
 ``On holographic thermalization,''
  Theor.\ Math.\ Phys.\  {\bf 174}, 186 (2013)
  arXiv:1211.6041 [hep-th]
  

 \bibitem{ABK} 
   I.~Aref'eva, A.~Bagrov and A.~S.~Koshelev,
  ``Holographic Thermalization from Kerr-AdS,''
  JHEP {\bf 1307}, 170 (2013)
  [arXiv:1305.3267 [hep-th]].

\bibitem{Li:2013sia} 
  Y.~-Z.~Li, S.~-F.~Wu, Y.~-Q.~Wang and G.~-H.~Yang,
  ``Linear growth of entanglement entropy in holographic thermalization captured by horizon interiors and mutual information,' 'JHEP {\bf 1309}, 057 (2013)
  [arXiv:1306.0210 [hep-th]].

\bibitem{Hartman:2013qma} 
  T.~Hartman and J.~Maldacena,
  ``Time Evolution of Entanglement Entropy from Black Hole Interiors,''
    JHEP {\bf 1305}, 014 (2013)
  [arXiv:1303.1080 [hep-th]].
 

 \bibitem{HL} 
  H.~Liu and S.~J.~Suh,
  ``Entanglement Tsunami: Universal Scaling in Holographic Thermalization,''
  Phys.\ Rev.\ Lett.\  {\bf 112}, 011601 (2014)
  [arXiv:1305.7244 [hep-th]].

 \bibitem{HLlong} 
  H.~Liu and S.~J.~Suh,
  ``Entanglement growth during thermalization in holographic systems,''
  Phys.\ Rev.\ D {\bf 89}, 066012 (2014) [arXiv:1311.1200 [hep-th]].
  
\bibitem{Leichenauer:2015xra} 
  S.~Leichenauer and M.~Moosa,
  ``Entanglement Tsunami in (1+1)-Dimensions,''
  Phys.\ Rev.\ D {\bf 92}, 126004 (2015)
  [arXiv:1505.04225 [hep-th]].






\bibitem{Shenker:2013pqa} 
  S.~H.~Shenker and D.~Stanford,
  ``Black holes and the butterfly effect,''   JHEP {\bf 1403}, 067 (2014)
  [arXiv:1306.0622 [hep-th]].
  
\bibitem{DAIA}  D.~S.~Ageev and I.~Y.~Aref'eva,
  ``Waking and Scrambling in Holographic Heating up,'' Theor. and Math. Phys., 193:1 (2017), 1534
  [arXiv:1701.07280 [hep-th]].
  
\bibitem{1312.6887} V. E. Hubeny and H. Maxfield, Holographic probes of collapsing black holes, JHEP 1403 (2014) 097 [arXiv:1312.6887[hep-th]]. 

\bibitem{Albash:2010mv} 
  T.~Albash and C.~V.~Johnson,
  ``Evolution of Holographic Entanglement Entropy after Thermal and Electromagnetic Quenches,''
  New J.\ Phys.\  {\bf 13}, 045017 (2011)
  [arXiv:1008.3027 [hep-th]].
\bibitem{Balasubramanian:2010ce} 
  V.~Balasubramanian, A.~Bernamonti, J.~de Boer, N.~Copland, B.~Craps, E.~Keski-Vakkuri, B.~Muller and A.~Schafer {\it et al.},
  ``Thermalization of Strongly Coupled Field Theories,''
  Phys.\ Rev.\ Lett.\  {\bf 106}, 191601 (2011)
  [arXiv:1012.4753 [hep-th]];

  
  
\bibitem{RT} 
  S.~Ryu and T.~Takayanagi,
  ``Holographic derivation of entanglement entropy from AdS/CFT,''
  Phys.\ Rev.\ Lett.\  {\bf 96}, 181602 (2006)
  [hep-th/0603001].

\bibitem{Hubeny:2007xt} 
  V.~E.~Hubeny, M.~Rangamani and T.~Takayanagi,
  ``A Covariant holographic entanglement entropy proposal,''
  JHEP {\bf 0707}, 062 (2007)
    [arXiv:0705.0016 [hep-th]].


\bibitem{Aparicio:2011zy} 
  J.~Aparicio and E.~Lopez,
  ``Evolution of Two-Point Functions from Holography,''
  JHEP {\bf 1112}, 082 (2011)
  [arXiv:1109.3571 [hep-th]].  
 
\bibitem{Allahbakhshi:2013rda} 
  D.~Allahbakhshi, M.~Alishahiha and A.~Naseh,
  ``Entanglement Thermodynamics,''
  JHEP {\bf 1308}, 102 (2013)
  [arXiv:1305.2728 [hep-th]].
 
\bibitem{Bhattacharya:2012mi} 
  J.~Bhattacharya, M.~Nozaki, T.~Takayanagi and T.~Ugajin,
  ``Thermodynamical Property of Entanglement Entropy for Excited States,''
  Phys.\  Rev.\  Lett.\  110, {\bf 091602} (2013)
  [arXiv:1212.1164 [hep-th]];
  
  M.~Nozaki, T.~Numasawa, A.~Prudenziati and T.~Takayanagi,
  ``Dynamics of Entanglement Entropy from Einstein Equation,''
  Phys.\ Rev.\ D {\bf 88}, no. 2, 026012 (2013)
  [arXiv:1304.7100 [hep-th]].

\bibitem{Blanco:2013joa} 
  D.~D.~Blanco, H.~Casini, L.~Y.~Hung and R.~C.~Myers,
  ``Relative Entropy and Holography,''
  JHEP {\bf 1308}, 060 (2013)
  doi:10.1007/JHEP08(2013)060
  [arXiv:1305.3182 [hep-th]].

\bibitem{Wong:2013gua} 
  G.~Wong, I.~Klich, L.~A.~Pando Zayas and D.~Vaman,
  ``Entanglement Temperature and Entanglement Entropy of Excited States,''
  JHEP {\bf 1312}, 020 (2013)
  [arXiv:1305.3291 [hep-th]].

\bibitem{Danielsson:1999zt}
  U.~H.~Danielsson, E.~Keski-Vakkuri and M.~Kruczenski,
  ``Spherically collapsing matter in AdS, holography, and shellons,''
  Nucl.\ Phys.\ B {\bf 563} (1999) 279
  [arXiv{hep-th/9905227}].

\bibitem{Erdmenger:2012xu}
  J.~Erdmenger and S.~Lin,
  ``Thermalization from gauge/gravity duality: Evolution of singularities in unequal time correlators,''
  JHEP {\bf 1210} (2012) 028
  [arxiv{arXiv:1205.6873} [hep-th]].
 




\bibitem{Galante:2012pv} 
   D.~Galante and M.~Schvellinger,
  ``Thermalization with a chemical potential from AdS spaces,''
  JHEP {\bf 1207}, 096 (2012)
  [arXiv:1205.1548 [hep-th]].



  \bibitem{Caceres:2012em}  
  E.~Caceres and A.~Kundu,
  ``Holographic Thermalization with Chemical Potential,''
  JHEP {\bf 1209}, 055 (2012)
  [arXiv:1205.2354 [hep-th]].
  
\bibitem{Baron}  
 W.~Baron, D.~Galante and M.~Schvellinger,
  ``Dynamics of holographic thermalization,''
  JHEP {\bf 1303}, 070 (2013)
  [arXiv:1212.5234 [hep-th]].
  

  



  \bibitem{Keranen}
  V.~Keranen, E.~Keski-Vakkuri and L.~Thorlacius,
  ``Thermalization and entanglement following a non-relativistic holographic quench,''
  Phys.\ Rev.\ D {\bf 85}, 026005 (2012)
  [arXiv:1110.5035 [hep-th]].

\bibitem{Alishahiha:2014cwa} 
  M.~Alishahiha, A.~Faraji Astaneh and M.~R.~Mohammadi Mozaffar,
  ``Thermalization in backgrounds with hyperscaling violating factor,''
  Phys.\ Rev.\ D {\bf 90}, no. 4, 046004 (2014)
  [arXiv:1401.2807 [hep-th]].

\bibitem{Fonda:2014ula} 
  P. Fonda, L. Franti, V. Keranen, E. Keski-Vakkuri, L. Thorlacius and E. Tonni,
  ''Holographic thermalization with Lifshitz scaling and hyperscaling violation,''
  JHEP {\bf 1408}, 051 (2014)
 
  [arXiv:1401.6088 [hep-th]].
  
     \bibitem{Arefeva:2015jkr} 
  I.~Ya.~Aref'eva,
  ``Formation time of quark-gluon plasma in heavy-ion collisions in the holographic shock wave model,''
Theor.\ Math.\ Phys.\  {\bf 184},  1239 (2015)
  [arXiv:1503.02185 [hep-th]].
    \bibitem{Aref'eva:2016dmy} 
  I.~Ya.~Aref'eva, A.~A.~Golubtsova and E.~Gourgoulhon,
  ``Analytic black branes in Lifshitz-like backgrounds and thermalization,''
  JHEP {\bf 1609}, 142 (2016)
  [arXiv:1601.06046 [hep-th]].




\bibitem{AV-photo} 
  I.~Aref'eva and I.~Volovich,
  ``Holographic Photosynthesis,''
  arXiv:1603.09107 [hep-th].
  

\bibitem{Landsteiner:2017lwm} 
  K.~Landsteiner, E.~Lopez and G.~Milans del Bosch,
  ``Quenching the CME via the gravitational anomaly and holography,''
  arXiv:1709.08384 [hep-th].



  
 \bibitem{Bhaseen:2013ypa} 
  M.~J.~Bhaseen, B.~Doyon, A.~Lucas and K.~Schalm,
  ``Far from equilibrium energy flow in quantum critical systems,''
  Nature Phys.\  {\bf 11}, 5 (2015)
  doi:10.1038/nphys3220
  [arXiv:1311.3655 [hep-th]].

\bibitem{Erdmenger:2017gdk} 
  J.~Erdmenger, D.~Fernandez, M.~Flory, E.~Megias, A.~K.~Straub and P.~Witkowski,
  ``Time evolution of entanglement for holographic steady state formation,''
  JHEP {\bf 1710}, 034 (2017)
  [arXiv:1705.04696 [hep-th]].
  

 \bibitem{Hubeny:2013hz} 
  V.~E.~Hubeny, M.~Rangamani and E.~Tonni,
  ``Thermalization of Causal Holographic Information,''
    JHEP {\bf 1305}, 136 (2013)
  [arXiv:1302.0853 [hep-th]].
 
\bibitem{Ziogas:2015aja} 
  V.~Ziogas,
  ``Holographic mutual information in global Vaidya-BTZ spacetime,''
  JHEP {\bf 1509}, 114 (2015)
  [arXiv:1507.00306 [hep-th]].
\bibitem{1602.05934}
  S.~Kundu and J.~F.~Pedraza,
  ``Spread of entanglement for small subsystems in holographic CFTs,''
  Phys.\ Rev.\ D {\bf 95}, 086008 (2017)
  [arXiv:1602.05934 [hep-th]].
 \bibitem{Calabrese:2005in} 
  P.~Calabrese and J.~L.~Cardy,
  ``Evolution of entanglement entropy in one-dimensional systems,''
  J.\ Stat.\ Mech.\  {\bf 0504}, P04010 (2005)
  [cond-mat/0503393].
\bibitem{Calabrese:2006} 
 P.~Calabrese and J.~Cardy,
``Time-dependence of correlation functions following a quantum quench,''
Phys. \ Rev. \ Lett. {\bf 96} 13680 (2006); arXiv:cond-mat/0601225.
\bibitem{QQT} S. Sotiriadis, P. Calabrese and J. Cardy, "Quantum Quench from a Thermal Initial State", EPL (2009) 20002 [arXiv:0903.0895 [cond-mat]]

\bibitem{Mezei1} 
  H.~Casini, H.~Liu and M.~Mezei,
  ``Spread of entanglement and causality,''
  JHEP {\bf 1607}, 077 (2016)
   [arXiv:1509.05044 [hep-th]].

\bibitem{Mezei2} 
  M.~Mezei,
  ``On entanglement spreading from holography,''
  JHEP {\bf 1705}, 064 (2017)
   [arXiv:1612.00082 [hep-th]].


\bibitem{Cardy:2015xaa} 
  J.~Cardy,
  ``Quantum Quenches to a Critical Point in One Dimension: some further results,''
  J.\ Stat.\ Mech.\  {\bf 1602}, no. 2, 023103 (2016)
    [arXiv:1507.07266 [cond-mat.stat-mech]].

\bibitem{Balasubramanian:2013oga} 
  V.~Balasubramanian {\it et al.},
  ``Inhomogeneous holographic thermalization,''
  JHEP {\bf 1310}, 082 (2013)
  [arXiv:1307.7086 [hep-th]].

\bibitem{Verlinde} 
  L.~McGough, M.~Mezei and H.~Verlinde,
  ``Moving the CFT into the bulk with $T\bar T$,''
  arXiv:1611.03470 [hep-th].
  

  \bibitem{AKT} 
  I.~Y.~Aref'eva, M.~A.~Khramtsov and M.~D.~Tikhanovskaya,
  ``Thermalization after holographic bilocal quench,''
  JHEP {\bf 1709}, 115 (2017)
  [arXiv:1706.07390 [hep-th]].

\bibitem{Nozaki:2013wia} 
  M.~Nozaki, T.~Numasawa and T.~Takayanagi,
  ``Holographic Local Quenches and Entanglement Density,''
  JHEP {\bf 1305}, 080 (2013)
  [arXiv:1302.5703 [hep-th]].

\end{thebibliography}
\end{document}